\documentclass[12pt]{article}
\usepackage{latexsym}
\usepackage{amsmath,calrsfs}
\usepackage{amsfonts}
\usepackage{amssymb}
\usepackage{amscd}
\usepackage{bbm}
\usepackage{fancybox}
\usepackage{cite}
\usepackage{amsmath,amsfonts,amsbsy}
\usepackage{pstricks,pst-node}
\usepackage[small,bf,hang]{caption2}
\usepackage{graphicx,xcolor}
\usepackage{epsfig}
\usepackage{psfrag}
\usepackage{comment}

\usepackage{float}

\psset{unit=1.3cm,linewidth=.5pt,radius=.2}  

\usepackage{multirow}                     
\usepackage{float}                          
\usepackage{lscape}                         
\usepackage{bm}

\newcommand{\beq}{\begin{equation}}
\newcommand{\eeq}{\end{equation}}
\newcommand{\bt}{\begin{tabular}}
\newcommand{\et}{\end{tabular}}
\newcommand{\bc}{\begin{center}}
\newcommand{\ec}{\end{center}}

\newcommand{\be}{\begin{equation}}
\newcommand{\ee}{\end{equation}}
\newcommand{\bea}{\begin{eqnarray}}
\newcommand{\eea}{\end{eqnarray}}
\newcommand{\ba}{\begin{array}}
\newcommand{\ea}{\end{array}}
\newcommand{\HH}{\mathcal{H}}

\def\bbox{{\,\lower0.9pt\vbox{\hrule \hbox{\vrule height 0.2 cm
\hskip 0.2 cm \vrule height 0.2 cm}\hrule}\,}}
\newcommand{\dsl}{\pa \kern-0.5em /}

\def\HH{\mathcal{H}}
\def\LL{\mathcal{L}}

\def\TT{\mathcal{T}}

\makeatletter \@addtoreset{equation}{section} \makeatother

\def\slashchar#1{\setbox0=\hbox{$#1$}           
   \dimen0=\wd0                                 
   \setbox1=\hbox{/} \dimen1=\wd1               
   \ifdim\dimen0>\dimen1                        
      \rlap{\hbox to \dimen0{\hfil/\hfil}}      
      #1                                        
   \else                                        
      \rlap{\hbox to \dimen1{\hfil$#1$\hfil}}   
      /                                         
   \fi}

\date{}

\begin{document}

\begin{titlepage}

\begin{center}

\vskip 1.5cm

{\Large \bf Simplified self-dual electrodynamics}
\smallskip

\vskip 1cm

{\bf  Jorge G.~Russo\,${}^{a,b}$ and  Paul K.~Townsend\,${}^c$} \\

\vskip 25pt

{\em $^a$  \hskip -.1truecm
\em Instituci\'o Catalana de Recerca i Estudis Avan\c{c}ats (ICREA),\\
Pg. Lluis Companys, 23, 08010 Barcelona, Spain.
 \vskip 5pt }

\vskip .4truecm

{\em $^b$  \hskip -.1truecm
\em Departament de F\' \i sica Cu\' antica i Astrof\'\i sica and Institut de Ci\`encies del Cosmos,\\ 
Universitat de Barcelona, Mart\'i Franqu\`es, 1, 08028
Barcelona, Spain.
 \vskip 5pt }
 
 \vskip .4truecm

{\em $^c$ \hskip -.1truecm
\em  Department of Applied Mathematics and Theoretical Physics,\\ Centre for Mathematical Sciences, University of Cambridge,\\
Wilberforce Road, Cambridge, CB3 0WA, U.K.\vskip 5pt }

\hskip 1cm

\noindent {\it e-mail:}  {\texttt  jorge.russo@icrea.cat, pkt10@cam.ac.uk}

\end{center}

\vskip 0.5cm
\begin{center} {\bf ABSTRACT}\\[3ex]
\end{center}

We present a new formulation of self-dual nonlinear electrodynamics in which interactions are determined by an auxiliary-field potential, with causality ensuring a unique solution to the auxiliary-field equation. The long-standing problem of an explicit Lagrangian for the generic `analytic' theory is simply solved by restriction to potentials that are even functions of the auxiliary field. In this case the Lagrangian can be linearised in quadratic field-strength scalars by the introduction of an additional pseudoscalar auxiliary field; this generalises, to all analytic self-dual theories, a well-known construction of the Born-Infeld theory.


\vfill

\end{titlepage}
\tableofcontents

\section{Introduction}
\setcounter{equation}{0}

Generic theories of nonlinear electrodynamics (NLED) have both a Lagrangian and a Hamiltonian formulation. 
The Lagrangian formulation allows Lorentz invariance to be made manifest because the standard Lagrangian density 
is a function $\LL(S,P)$ of the Lorentz (pseudo)scalars $(S,P)$ that are quadratic in the 
2-form field-strength $F=dA$ for 1-form potential $A$ on 4D Minkowski spacetime. In terms of the electric and 
magnetic field components $({\bf E},{\bf B})$ they are (in cartesian coordinates) 
\be
S :=-\frac14 F_{\mu\nu}F^{\mu\nu}= \frac12\left(|{\bf E}|^2 - |{\bf B}|^2\right)\, , \qquad P :=-\frac14 F_{\mu\nu}\tilde F^{\mu\nu}= {\bf E}\cdot{\bf B}\, . 
\ee
The first example was Born's 1933 model:  $\LL_{\rm Born} = T-\sqrt{T^2-2TS}$, 
where the constant $T$ has dimensions of energy density; it 
determines a ``Born scale''. It was soon followed by the Born-Infeld (BI) model: 
\be\label{LBI}
\LL_{\rm BI} = T- \sqrt{T^2-2TS-P^2}\, .
\ee

The Hamiltonian density is the Legendre transform of $\LL$ with respect to ${\bf E}$:
\be\label{LtoH}
\HH({\bf D}, {\bf B}) = \sup_{{\bf E}}\left\{ {\bf E} \cdot {\bf D} - \LL({\bf E},{\bf B})\right\}\, . 
\ee
For the Born model one finds that $\HH_{\rm Born} = \sqrt{(T+ |{\bf D}|^2) (T+ |{\bf B}|^2)} $, which  is 
invariant under the $Z_2$ electromagnetic duality transformation ${\bf D} \leftrightarrow {\bf B}$.
For BI one finds instead that 
\be\label{HBI}
\HH_{\rm BI} = \sqrt{T^2 + 2Ts + p^2} -T\, , 
\ee
where
\be
s= \frac12\left(|{\bf D}|^2 + |{\bf B}|^2\right)\, , \qquad p= |{\bf D}\times {\bf B}|\, . 
\ee
These two rotation (and parity) invariants are also invariant under a  $U(1)$ electromagnetic duality transformation acting on the complex 3-vector density ${\bf D} +i{\bf B}$ by a phase shift, and any function $\HH(s,p)$ has the same property\footnote{Generically, $\HH$ can be expressed as a function of 
$\frac12|{\bf D}|^2$, $\frac12 |{\bf B}|^2$ and ${\bf D}\cdot {\bf B}$.}. We shall say that any NLED with a Hamiltonian function $\HH(s,p)$ is ``self-dual''; BI is a simple example. Thus, the Hamiltonian formulation allows self-duality to be made manifest at the cost of a loss of manifest Lorentz invariance. We refer the reader to \cite{Russo:2024kto} for more details, and citations prior to the revival of interest in this topic that began around 1980, with the supergravity-inspired work on electromagnetic duality by Gaillard and Zumino  \cite{Gaillard:1981rj}.

The main aim of this paper is to present a novel reformulation of the generic Lorentz invariant and self-dual NLED. 
 It involves the introduction of a single scalar auxiliary field $\phi$ with a potential function $W(\phi)$. Consider the following Lagrangian density:
\be\label{bbb1}
\boxed{\LL(S,P;\phi) = (\cosh\phi)  S + (\sinh\phi)  \sqrt{S^2+P^2} \ -\  W(\phi)} 
\ee
The auxiliary-field equation is 
\be\label{Laux}
W'(\phi) =  (\cosh\phi)  \sqrt{S^2+P^2} + (\sinh\phi)S \, . 
\ee
Notice that $W'\ge0$ with equality for $S=P=0$, which includes the vacuum. The `vacuum value' $\phi_0$ of $\phi$ is therefore found as the solution to 
\be\label{vacphi}
W'(\phi_0) =0\, .
\ee
We shall assume, for the moment, that this solution exists and is unique, and that this remains true of  \eqref{Laux} for sufficiently small $|S|$ and $|P|$ because this is required for the existence of a weak-field expansion; we shall see later that uniqueness  is implied by causality. For any function $W$ restricted in this way, we claim that the Lagrangian density of \eqref{bbb1} defines a self-dual NLED (with a weak-field expansion).  For example, 
the Born-Infeld theory corresponds to the choice $W(\phi)=2T\sinh^2(\phi/2)$,  as may easily be verified. 

There are various ways in which our claim can be verified in general.  One is to verify that the Lagrangian density $\LL(S,P)$ resulting from the elimination of the auxiliary field satisfies a condition on it that is known to be required for self-duality. We will do this in the following section because it will be needed
to establish completeness of the new formulation (i.e. that no cases are left out),  but there is another way to verify our claim that is in some respects 
simpler\footnote{Alternatively, we could promote the ModMax parameter $\gamma$ to a scalar field $\phi$ and then add the potential term. We thank Sergei Kuzenko for the observation, based on results for supersymmetric NLED in \cite{Kuzenko:2000tg,Kuzenko:2000uh}, that this process preserves self-duality, as our results to follow confirm.}. It is based on the observation that any NLED defined by $\LL(S,P)$ is self-dual iff its corresponding Hamiltonian density can be expressed as a function of $(s,p)$ only\footnote{This assumes a normalisation for the self-dual 
$\LL$ equivalent to a choice of units for which the speed of light is unity, but this choice may be made without loss of generality.}. For the Lagrangian density of  \eqref{bbb1} the corresponding Hamiltonian density is
\be
\HH({\bf D},{\bf B}; \phi)  = \sup_{{\bf E}}\left\{ {\bf E} \cdot {\bf D}  - 
 \left[(\cosh\phi)  S + (\sinh\phi)  \sqrt{S^2+P^2}\right]\right\} + W(\phi)\, ,  
\ee 
and the $\phi$ field equation is now $\partial\HH/\partial\phi=0$. The variational problem here is mathematically equivalent to the problem, solved in \cite{Bandos:2020jsw}, of finding the Hamiltonian density of ModMax electrodynamics from its Lagrangian density. The result in the current context is
\be\label{ccc1}
\boxed{\HH({\bf D},{\bf B};\phi) =  (\cosh\phi) s - (\sinh\phi)  
\sqrt{s^2-p^2} \ + \ W(\phi)} 
\ee
but with the following restriction\footnote{See
\cite{Bandos:2020jsw, Russo:2024ptw} and especially \cite{Russo:2025wph} for a detailed discussion of this restriction and its implications.}  on the domain of $\HH$: 
\be\label{restrict}
s\ge(\cosh\phi)p\, . 
\ee 
The equation for $\phi$ is now
\be\label{Haux}
W'(\phi) =  (\cosh\phi)\sqrt{s^2-p^2} -(\sinh\phi) s\, , 
\ee
and back-substitution yields a Hamiltonian density $\HH(s,p)$. Because of the restriction \eqref{restrict} it is again true that $W'\ge0$, with equality 
when $s=p=0$, which is the vacuum\footnote{In fact, $s=0$ implies $|{\bf D}|=|{\bf B}|=0$, and hence $p=0$.}.

The boxed equations \eqref{bbb1} and \eqref{ccc1} therefore establish a correspondence between a class of self-dual Lorentz-invariant NLED theories and a class of functions $W(\phi)$. The precise nature of these classes has still to be clarified, but let us first illustrate the correspondence with a simple example that includes Born-Infeld as a special case:  
\be\label{WMMB}
W_{\rm MMB}(\phi) = T[\cosh(\phi-\gamma) -1]\, , 
\ee
where $\gamma$ is a dimensionless constant and $T$ determines a Born-scale; the `MMB' subscript will be explained below. In this case 
$\phi_0= \gamma$ since $W'(\gamma)=0$, and the constant term has been chosen such that $W(\phi_0) =0$; i.e. such that the vacuum energy is zero.  
This potential function leads to the Lagrangian density
\be\label{LMMB}
\LL_{\rm MMB}(S,P) =T -\sqrt{T^2-2T \LL_{\rm MM}(S,P)-P^2}\, , 
\ee
where
\be
\LL_{\rm MM}(S,P)=(\cosh\gamma)\, S+(\sinh\gamma)\,  \sqrt{S^2+P^2}\, . 
\ee
Notice that the weak-field limit found by taking $T\to\infty$ is $\LL_{\rm MM}$, which is the most general conformal weak-field limit for a self-dual NLED, with $\gamma\ge0$ required for causality \cite{Bandos:2020jsw}; Maxwell is the $\gamma=0$ case and ``ModMax'' (an abbreviation for modified Maxwell) is the $\gamma>0$ case. The non-conformal NLED defined by \eqref{LMMB} for $\gamma=0$ is Born-Infeld. For $\gamma>0$ it is the generalisation of 
Born-Infeld introduced in \cite{Bandos:2020jsw,Bandos:2020hgy} that (for brevity) we call ``ModMaxBorn'' (MMB) because its weak-field limit is ModMax. The MMB Hamiltonian density can now be found from \eqref{ccc1} by using the same function $W(\phi)$; i.e. \eqref{WMMB}. This yields 
\be\label{HMMB}
\HH_{\rm MMB}(s,p) = \sqrt{T^2 + 2T\HH_{\rm MM} + p^2} -T\, , 
\ee
where $\HH_{\rm MM}$ is the ModMax Hamiltonian density:
\be
H_{\rm MM}(s,p) = (\cosh\gamma) s - (\sinh\gamma)\sqrt{s^2-p^2}\, .  
\ee

Assuming that the constant term in $W$ has been fixed by requiring zero vacuum energy, any non-zero choice of $W$  necessarily introduces a Born scale because $W$ has dimensions of energy density whereas its argument $\phi$ is dimensionless. This means that the only possible function $W$ that could yield a {\sl conformal} self-dual NLED is the zero function $W(\phi)\equiv 0$. When this is used in \eqref{ccc1} the auxiliary-field equation has the solution $e^\phi = \sqrt{v/u}$. Back-substitution then yields $\HH=p$, which is the Hamiltonian density for the conformal Bialynicki-Birula (BB) electrodynamics \cite{Bialynicki-Birula:1984daz}, originally found from the strong-field ($T\to0$) limit of the Born-Infeld Hamiltonian density \eqref{HBI}. When the same zero potential is used in \eqref{bbb1} the auxiliary-field equation \eqref{Laux} requires $S=P=0$, for any $\phi$. This is to be expected because there is no `standard' Lagrangian density for BB electrodynamics but its field equations are equivalent to $S=P=0$ \cite{Bialynicki-Birula:1992rcm}. 

A corollary of this discussion of the zero-potential case is that there is no choice of $W(\phi)$ for the conformal Maxwell/ModMax cases! Of course, they are included implicitly as weak-field limits but the weak-field limit must be taken {\sl after} elimination of the auxiliary field. The MMB potential function of \eqref{WMMB} illustrates this fact because it does not have a well-defined $T\to\infty$ limit. There is more that can be said on this point but it does not alter the fact that there is no function $W(\phi)$ for the Maxwell/ModMax cases consistent with the auxiliary field interpretation for $\phi$. This surprising feature makes our new auxiliary-field formulation essentially different from previous auxiliary-field formulations, e.g. that of Ivanov and Zupnik \cite{Ivanov:2003uj,Ivanov:2012bq,Ivanov:2013jba}. 

We should also mention here other formulations  of self-dual NLED theories in which additional fields are introduced to allow both Lorentz invariance and self-duality to be made simultaneously manifest (at the cost of additional non-manifest gauge invariances) e.g. \cite{Pasti:1995tn} and \cite{Avetisyan:2021heg} (see \cite{Sorokin:2025ezs} for a recent review of this approach).   However, our auxiliary-field formulation achieves something similar because it makes manifest the fact that the (manifestly Lorentz-invariant) Lagrangian and (manifestly self-dual) Hamiltonian are related by a simple algebraic ``substitution rule''. This is because a comparison of \eqref{bbb1} with \eqref{ccc1} shows that $\HH$, viewed as a function of $s$ and $\sqrt{s^2-p^2}$,  can be found from $-\LL$, viewed as a function of $S$ and $\sqrt{S^2+P^2}$, by the substitution 
\be
S \to  -s\, , \qquad \sqrt{S^2+P^2} \to \sqrt{s^2-p^2}\, ,  
\ee
and the inverse substitution takes $\HH$ to $-\LL$. This ``substitution rule'' was derived in \cite{Russo:2024ptw} by a much more circuitous route.

One advantage of our new auxiliary-field formulation of self-dual NLED is that it allows a simple answer to a question raised by Perry and Schwarz  in  the related context of their formulation of chiral 2-form electrodynamics \cite{Perry:1996mk} (see \cite{Bandos:2020hgy,Russo:2025wph} for details of this relation). It is obvious from \eqref{bbb1} that the weak-field expansion of $\LL(S,P)$ is generically an expansion in powers of $S$ and $\sqrt{S^2+P^2}$. However, there are cases for which it is an expansion in powers of $S$ and $P^2$; i.e. analytic  in the Lorentz scalars $(S,P)$ (and even in $P$ because $P$ is parity odd and $\LL$ is parity-even). One can see from \eqref{LBI} that Born-Infeld is an example, but is there some simple criterion that distinguishes self-dual theories with ``analytic'' weak-field expansions? A {\sl necessary} condition (illustrated by the MMB example) is that the weak-field limit should be Maxwell, but this is insufficient because it does not exclude weak-field expansions in which the non-analyticity appears at higher order; e.g. a term proportional to $S^2\sqrt{S^2+P^2}$ at the $S^3$ level; in fact, an explicit example with this term will be discussed in section \ref{sec:phi-parity}.

A criterion for analyticity was proposed by Perry and Schwarz \cite{Perry:1996mk}, and has been discussed
in the 4D self-dual NLED context by Bekaert and Cucu \cite{Bekaert:2001wa}. It is a condition that could be used on a case-by-case basis but it does not provide an explicit formula for the general analytic Lagrangian density. In contrast, the criterion for an analytic weak-field expansion in our new auxiliary-field formulation is both simple and simply implemented:  the expansion is analytic iff 
\be\label{phiparity}
W(\phi)= W(-\phi)\, . 
\ee
We shall say that a function with this property is even under ``$\phi$-parity'', or ``symmetric'', and that the corresponding self-dual NLED is ``$\phi$-parity self-dual'', or ``analytic'' because its weak-field expansion is analytic in $(S,P)$. The concept of $\phi$-parity thus defined is essentially the same as the ``$\Phi$-parity'' introduced in \cite{Russo:2024ptw} in a context without an auxiliary field, and this motivated our choice of notation for the auxiliary field. 

As observed above, the fact that the Lagrangian density of \eqref{bbb1} is linear in $S$ and 
$\sqrt{S^2+P^2}$ implies that the weak-field expansion is in powers of $S$ and $\sqrt{S^2+P^2}$. 
This suggests that for any {\sl analytic} self-dual NLED it should be possible to introduce
an additional auxiliary field to `linearise' $\LL$ in $S$ and $P$. This is indeed possible, and 
we present the Lagrangian density in section \ref{sec:RTgen}. Remarkably, in the particular case of BI, it is equivalent to the auxiliary-field formulation of BI introduced long ago by Ro{\v c}ek and Tseytlin \cite{Rocek:1997hi}. Thus, another result of this paper is a generalisation of the Ro{\v c}ek-Tseytlin 
auxiliary-field formulation of BI to {\sl all} analytic self-dual NLED. 

In the following two sections we shall establish the generality of our new auxiliary field formulation, present the conditions on $W$ required by causality, and show that these conditions imply that any solution of the auxiliary-field equation is unique. In section \ref{sec:CS} we review the topic of causality, and the derivation of causality conditions for self-dual NLED, with many simplifications. In section \ref{sec:phi-parity} we explain why an analytic weak-field expansion requires a symmetric potential $W$, and in section \ref{sec:RTgen} we present the generalisation to all analytic self-dual NLED of the Ro{\v c}ek-Tseytlin formulation of the Born-Infeld theory. In section \ref{sec:energy} we revisit the topic of energy conditions on the stress-energy tensor of NLED theories to show how these are implied by the causality conditions on $W$.

\section{Preliminaries: new variables}\label{sec:prelim}

The starting point for most investigations of general NLED theories has been a Lagrangian density function $\LL(S,P)$, and the condition for any such function to define a self-dual NLED was first obtained by Bialynicki-Birula in a 1983 review of NLED based  on the Macroscopic Maxwell 
equations. We recall that the {\sl source-free} equations are
\be\label{MacroMax}
\begin{aligned}
\dot{\bf B} &= -\boldsymbol{\nabla} \times {\bf E}\,, \qquad 
\boldsymbol{\nabla}\cdot {\bf B}=0 \\
\dot{\bf D} &= \boldsymbol{\nabla} \times {\bf H}\, ,\qquad \ \
\boldsymbol{\nabla}\cdot {\bf D}=0\, , 
\end{aligned}
\ee
taken together with “constitutive relations” between the 3-covector fields
$({\bf E},{\bf H})$ and the 3-vector densities $({\bf D},{\bf B})$. These 
can be expressed in various ways; e.g. via the choice of a Hamiltonian density $\HH({\bf D},{\bf B})$ or Lagrangian density $\LL({\bf E},{\bf B})$, in which case 
\be
({\bf E}, {\bf H}) = \left(\frac{\partial \HH}{\partial{\bf D}},  
\frac{\partial \HH}{\partial{\bf B}}\right) \, , \qquad 
({\bf D},{\bf H})  = \left(\frac{\partial \LL}{\partial{\bf E}},  
-\frac{\partial \LL}{\partial{\bf B}}\right)\,. 
\ee
We have written the equations \eqref{MacroMax} in standard vector calculus notation for cartesian coordinates but they are coordinate independent. The 3-metric appears only in the choice of $\HH$, which determines $\LL$ by the inverse of \eqref{LtoH}:
\be\label{HtoL}
\LL({\bf E},{\bf B}) = \sup_{\bf D} \left\{ {\bf D}\cdot{\bf E} - \HH({\bf D},{\bf B})\right\}\, .
\ee
For standard Euclidean coordinates, and a choice of units such that the {\it in vacuo} speed of light is unity, the free-field Maxwell theory corresponds to $\HH =s$, which yields $\LL=S$. 

In this context, the  respective conditions for any NLED to be Lorentz invariant and self-dual are \cite{Bialynicki-Birula:1984daz}
\be\label{LI+SD}
{\bf D}\times {\bf B}= {\bf E} \times {\bf H}\, , \qquad {\bf E}\cdot{\bf B} = {\bf D}\cdot{\bf H}\, . 
\ee
The first of these conditions is an identity for any Lagrangian density $\LL(S,P)$ but the second requires $\LL(S,P)$ to solve a particular partial differential equation (PDE) \cite{Bialynicki-Birula:1984daz}. This PDE takes the much simpler form found by Gibbons and Rasheed \cite{Gibbons:1995cv} once the function $\LL$ is expressed in terms of new independent variables (see below). Conversely, the self-duality condition is an identity for any Hamiltonian density $\HH(s,p)$, but the Lorentz-invariance condition now requires $\HH(s,p)$ to solve a particular PDE, and this too can be similarly simplified by the introduction of new independent variables. The details are as follows:

\begin{itemize}

\item For $\LL(S,P)$ we introduce new Lorentz scalars $(U,V)$ by setting 
\be
S=V-U \, , \qquad P^2= 4UV\, , 
\ee
 where $UV\ge0$ necessarily, and we may choose both $U$ and $V$ to be non-negative. The condition for $\LL(U,V)$ to define a self-dual NLED 
 is very simple \cite{Gibbons:1995cv}:
 \be\label{PDEL}
 \LL_U \LL_V =-1\, . 
 \ee

\item For $\HH(s,p)$ we introduce new duality-invariant rotation scalars by setting 
 \be
 s= v+u \, , \qquad p= 2\sqrt{uv}\, , 
 \ee
 where $u$ and $v$ are non-negative necessarily and we may choose\footnote{This is opposite to the choice made in \cite{Bandos:2020jsw,Bandos:2020hgy}, where $u\ge v$.} $v\ge u$. The condition for $\HH(u,v)$ to define a Lorentz invariant NLED is \cite{Bandos:2020hgy}
\be\label{PDEH}
\HH_u\HH_v =1\, .
\ee
This is formally equivalent to \eqref{PDEL} since the sign on the right-hand side can be changed by a sign change for one of the two independent 
variables, but then one variable would be non-positive rather than non-negative. 

\end{itemize}

Using these variables it is easy to confirm that both \eqref{bbb1} and \eqref{ccc1} define a self-dual Lorentz-invariant NLED, as follows: 

\begin{itemize}

\item We first rewrite \eqref{bbb1} as 
 \be\label{bbb2}
 \LL(U,V;\phi) = e^\phi V - e^{-\phi}U - W(\phi)\, .   
 \ee
 Taking the exterior derivative on both sides we get
 \be
 d\LL = e^\phi dV - e^{-\phi}dU + \frac{\partial\LL}{\partial\phi} d\phi\, . 
\ee
If $\phi=\phi(U,V)$ is the solution to $\partial\LL/\partial\phi=0$ then the function $\LL(U,V)$ found by elimination of $\phi$ has the partial derivatives
\be
\LL_V= e^{\phi(U,V)} \, , \qquad \LL_V= - e^{-\phi(U,V)} \quad \Rightarrow \quad \LL_U\LL_V=-1\, . 
\ee
This confirms our claim that \eqref{bbb1} defines a self-dual NLED.

\item We first rewrite \eqref{ccc1} as
\be\label{ccc2} 
\HH(u,v;\phi) = e^{-\phi}v + e^\phi u + W(\phi)\, . 
\ee
If $\phi=\phi(u,v)$ is the solution of $\partial \HH/\partial\phi =0$ then elimination of $\phi$ yields a function $\HH(u,v)$ with 
\be
\HH_u= e^{\phi(u,v)}\, , \qquad \HH_v = e^{-\phi(u,v)} \quad 
\Rightarrow \quad \HH_u\HH_v =1\, . 
\ee
This confirms that \eqref{ccc1} defines a Lorentz-invariant NLED. 

\end{itemize}
The fact that \eqref{bbb2} and \eqref{ccc2} define the {\sl same} NLED
theory relies on the fact that one is the Legendre transform of the other, 
but this transform maps {\sl linear} functions in $(U,V)$ to {\sl linear} 
functions in $(u,v)$, and {\sl vice-versa}, and once this map is known 
it determines the relation between $\LL(U,V)$ and $\HH(u,v)$ for generic Lorentz-invariant 
self-dual NLED. 

We may now restate the ``substitution rule'' that takes the Lagrangian to the Hamiltonian or {\it vice versa} as follows. Starting with \eqref{bbb2} we change the overall sign and then replace $(U,V)$ by $(v,-u)$ to get \eqref{ccc2}. Similarly, starting with \eqref{ccc2} we change the overall sign and then replace $(u,v)$ by $(-V,U)$. In other words, we exchange (simultaneously) the dependent and independent variables as follows: 
\be\label{subrule}
\LL \leftrightarrow -\HH\, ;  \qquad U\leftrightarrow v\, , \quad V \leftrightarrow -u \, . 
\ee
This can be illustrated by the MMB example of the introduction. Starting from \eqref{bbb1} with the function $W$ of \eqref{WMMB}, the $\phi$ equation is solved by
\be
\label{solphi}
e^{\phi(U,V)} = e^\gamma \sqrt{\frac{T+2e^{-\gamma}U}{T-2e^\gamma V}}\, ,   
\ee
and this yields
\be\label{LLMMB}
\LL_{\rm MBB}(U,V) = T-\sqrt{(T+2e^{-\gamma}U)(T-2e^\gamma V)}\, ,
\ee
which is equivalent to \eqref{LMMB}. From this result we can use \eqref{subrule} to deduce that 
\be
\HH_{\rm MMB}(u,v) = \sqrt{(T+2e^{-\gamma}v)(T+2e^\gamma u)} -T\, ,  
\ee 
which is equivalent to \eqref{HMMB}.

\section{Generality and Causality}\label{sec:GC}

It was observed in \cite{Gibbons:1995cv} that the possible weak-field expansions of solutions of \eqref{PDEL} depend on the choice of a single function of one variable. In fact, the general solution in terms of the boundary function $\LL(0,V)=\ell(V)$ can be 
found in the text by Courant and Hilbert \cite{C&H}; in the current context it is \cite{Gaillard:1997rt,Russo:2024llm}
\be\label{CH-Lag}
\LL(U,V) = \ell(\tau) -\frac{2U}{\dot\ell(\tau)}\, , \qquad 
\tau= V+ \frac{U}{[\dot\ell(\tau)]^2}\, \qquad (\dot\ell >0). 
\ee
Notice, from its definition, that $\tau$ is non-negative and has the dimensions of energy density, as does the ``CH-function'' $\ell$. 
This means that the only functions $\ell(\tau)$ that do not introduce a Born scale are linear functions, which define conformal NLED theories. 

If we insist on a weak-field expansion about a conformal theory with non-negative energy, e.g. Maxwell, then $\ell(\tau)$ must have a Taylor expansion of the form 
\be
\ell(\tau) = e^\gamma \tau + \mathcal{O}(\tau^2)\, ,
\ee
where $\gamma$ is a dimensionless constant; a possible constant term is omitted because we assume zero vacuum energy. The main result of\cite{Russo:2024llm} was that causality (no superluminal wave perturbations of any constant uniform electromagnetic background) imposes the following conditions on $\ell(\tau)$ 
\be\label{causcon}
\dot\ell(\tau) \ge 1\, , \qquad \ddot \ell(\tau)\ge 0  \, . 
\ee
We postpone to the following section an explanation of the origin and significance of these two conditions. The first of them requires $\gamma\ge0$, which means that the weak-field limit is either Maxwell ($\gamma=0$) or ModMax ($\gamma>0$), as illustrated by the MMB example of the previous sections.  
The second condition is just the statement that $\ell(\tau)$ must be a convex function. In the Maxwell/ModMax cases $\ell(\tau)$ is convex but not 
``strictly convex'' since $\ddot\ell=0$. For all other cases $\ell(\tau)$
is strictly convex, i.e. $\ddot\ell(\tau)>0$ for $\tau>0$ (since $\ddot\ell(0)=0$ is not excluded) and this implies the existence of a Born scale.  

The convexity condition on $\ell(\tau)$ implies that it can be expressed as the Legendre-transform of some other convex function $\Omega(y)$: 
\be\label{ell-LT}
\ell(\tau) = \sup_y\left\{ y\tau  -\Omega(y)\right\} \, .
\ee
Since the Legendre transform is an involution when acting on convex functions, we also have 
\be\label{Om-LT}
\Omega(y) = \sup_\tau\left\{\tau y - \ell(\tau)\right\}\, .
\ee
The point of this observation is that we should be able to rewrite \eqref{CH-Lag} in a new form in which self-dual NLED Lagrangians are parameterised by functions $\Omega(y)$ rather than $\ell(\tau)$ since each of these functions uniquely determines the other. 

To proceed, we now suppose that $\ell(\tau)$ is not just convex but ``strictly convex''. The same will then be true of $\Omega(y)$, and we may rewrite \eqref{ell-LT} as 
\be\label{ell-Om}
\ell(\tau) = \tau y- \Omega(y) \, , \qquad \tau= \Omega'(y)\, , 
\ee
where the second equation is to be solved for $y=y(\tau)$; this solution is unique if it exists because $\Omega'(y)$ is an increasing function of $y$. Notice that 
\be
\dot\ell(\tau) = y \, , 
\ee
and hence 
\be\label{dderivs}
\ddot \ell(\tau) = dy/d\tau = 1/\Omega'{}'(y)\, , 
\ee
which confirms that $\Omega$ is strictly convex if $\ell$ is (and {\it vice versa}). In addition, we may now write the equation 
for $\tau$ in \eqref{CH-Lag} as $\tau=V+ U/y^2$. Using this in \eqref{ell-Om} and the result for $\ell(\tau)$ in \eqref{CH-Lag} we have 
\be\label{CH-alt}
\mathcal{L} = yV- \frac{U}{y} - \Omega(y)\, , \qquad \Omega'(y) = V +\frac{U}{y^2}\, \qquad (y>0).
\ee
The causality relations \eqref{causcon} can now be rewritten (taking into account the {\sl strict} convexity of $\ell$) as
\be\label{causalO}
y\ge1\, , \qquad  \Omega''(y)>0\, . 
\ee
For linear convex functions $\ell$ there is still a convex function $\Omega$ defined by \eqref{Om-LT} 
but it is defined only for one value of $y$; specifically, $y=e^\gamma$, and this yields the Maxwell/ModMax
cases \cite{Russo:2024ptw}. However, for the auxiliary-field formulation that we introduce in the following subsection, 
we must exclude these cases.  

To illustrate the conversion of a (nonlinear) CH-function $\ell$ to the corresponding $\Omega$-function, we shall consider the first few terms in the weak-field expansion of $\ell$:
\be
\ell(\tau) = T\left\{ e^\gamma \left(\frac{\tau}{T}\right) + \frac{a}{2}\left(\frac{\tau}{T}\right)^2
+ \mathcal{O}\left[\left(\frac{\tau}{T}\right)^3\right] \right\}\, , 
\ee
where $T$ is the Born scale. Assuming that $a>0$ to ensure strict convexity\footnote{To allow for $\ddot\ell(0)=0$ we would need to specify higher-order terms in the expansion.} the Legendre transform yields 
\be\label{exp-Om}
\Omega(y) = T \left\{ \frac{1}{2a} \left(y - e^\gamma\right)^2  + 
\mathcal{O}\left[ \left(y-e^\gamma\right)^3\right]\right\}\, . 
\ee

\subsection{The auxiliary field formulation}

In \cite{Russo:2024ptw} we arrived at \eqref{CH-alt} (using a slightly different notation\footnote{The variable $y$ and the function $\Omega(y)$ correspond to $1/x$ and $-\omega(x)$ in \cite{Russo:2024ptw}.}) from another solution of Courant and Hilbert for the PDE of \eqref{PDEL}; we therefore called it the ``alternative CH-construction''. What we failed to notice\footnote{Neither is it mentioned by Courant and Hilbert.} is that the second equation in \eqref{CH-alt} is the equation $\partial\LL/\partial y=0$, which allows the variable $y$ to be considered as an auxiliary field, albeit one subject to the constraint $y>0$. Thus, \eqref{CH-alt} can be replaced by 
\be\label{LUVy}
\LL(U,V;y) = yV- \frac{U}{y} - \Omega(y)\, \qquad (y>0)\, ,  
\ee
with $y$ being determined by its own field equation.  By defining 
\be\label{WOmy}
W(\phi) = \Omega(y)\, , \qquad y=e^\phi\, , 
\ee
we now have an unconstrained auxiliary scalar field $\phi$ with potential function $W$, and the Lagrangian density of \eqref{bbb2}, which is equivalent to \eqref{bbb1} of the Introduction.  Of course, both $\phi$ and $W(\phi)$ are constrained by causality; the conditions of \eqref{causalO} now become
\be\label{causal0}
\phi\ge0 \, , \qquad W''(\phi) > W'(\phi) \, ,   
\ee
where it should be understood that $\phi=\phi(U,V)$ solves the auxiliary-field equation
\be\label{Wp}
W'(\phi) = e^\phi V + e^{-\phi} U\, . 
\ee
Notice that the causality condition on $W$ implies that $W'(\phi)$ is a monotonically increasing function of $\phi$, which implies that $\phi(U,V)$ is a non-decreasing function of $W'(\phi)$ and hence that 
\be
\phi(U,V) \ge \phi_0\, , 
\ee
where we recall that  $\phi_0 \equiv \phi(0,0)$ is  the ``vacuum value'' of $\phi(U,V)$. In the MMB case, for example,  $\phi_0=\gamma$ and the solution for $\phi=\phi(U,V)$ of  \eqref{solphi} shows that $\phi\ge\gamma$, as expected because the condition $W''>W'$ is satisfied for all $\phi$. 
We may now rewrite the causality conditions of \eqref{causal1} more precisely as 
\be\label{causal1}
\phi_0 \ge0 \, , \qquad  W''(\phi) >W'(\phi) \qquad (\phi\ge \phi_0)\, . 
\ee

Returning to the expansion of $\Omega(y)$ in \eqref{exp-Om}, we may use \eqref{WOmy}  to find an equivalent expansion of $W(\phi)$ in powers of $(e^\phi- e^\gamma) = e^\gamma[e^{(\phi-\gamma)}-1]$, 
which can be rewritten as the following expansion in powers of $(\phi-\gamma)$:
\be\label{Wexpand}
W(\phi) = T\left\{ \frac{e^{2\gamma}}{2a} (\phi-\gamma)^2 + 
\mathcal{O}[ (\phi-\gamma)^3] \right\}\, . 
\ee 
The MMB potential function \eqref{WMMB} is an example with $a=e^{2\gamma}$, but we now see more generally that there is no $T\to\infty$ limit of $W(\phi)$.

We now explain why the causality conditions on $W$ imply that any
solution of the auxiliary-field equation of \eqref{Laux} is unique. This equation is  equivalent to the equation for the variable $y$ in \eqref{CH-alt}: 
\be\label{Aux-y}
\Omega'(y) = V+ \frac{U}{y^2}\, .  
\ee
The right-hand side is either constant (for $U=0$) or a monotonically decreasing function of $y$ (for $U>0$). Since $\Omega$ is strictly convex, given causality, the left-hand side is a monotonically increasing function of $y$. The graphs of these two functions can therefore intersect at most 
once. For values of $(U,V)$ for which they do not intersect the Lagrangian density $\LL(U,V)$ is not defined. This was to be expected; for example, it can be seen from \eqref{LLMMB} that $\LL_{\rm MMB}(U,V)$ is not defined for $2e^\gamma V> T$. The same arguments applies in the Hamiltonian formulation of \eqref{ccc1} because the auxiliary field equation is then equivalent to $\Omega'(y) = -u + v/y^2$. One difference is that there is now a solution for $\Omega(y)\equiv0$, which yields the conformal BB electrodynamics mentioned in the Introduction.

To summarise: we have now rederived the auxiliary-field formulation 
of the Introduction from a general parametrisation of the Lagrangians of self-dual NLED in terms of a CH-function that specifies the boundary conditions used to integrate the self-duality PDE. This result shows that the auxiliary-field formulation is general because it excludes only those conformal cases that are implicitly included 
as weak-field limits of the non-conformal cases.

\section{Causality simplified}\label{sec:CS}
\setcounter{equation}{0}

In the previous section we have used extensively the causality conditions 
of \eqref{causcon} for self-dual NLED, and re-writings of them such as \eqref{causalO} and \eqref{causal1}. Here we provide a brief overview of the topic, including a conceptually simpler route to the causality conditions of \eqref{causcon} for self-dual NLED. 

Conditions for causality of generic NLED theories defined by a Lagrangian density 
$\LL(S,P)$ were obtained by Boillat and Plebanski in the 1960s by considering shock waves (i.e. propagating discontinuities) in generic smooth background electromagnetic fields (see \cite{DeLorenci:2000yh} for a more recent account of this method). Equivalent results are found by linearisation of the NLED field equations about stationary homogeneous electromagnetic background, which can be viewed as optical media in which perturbations propagate  \cite{Bialynicki-Birula:1984daz} (see \cite{Russo:2022qvz} for a more recent account of this method). Either way, one finds propagation of two polarisation modes, generically with different dispersion relations; this is the phenomenon of birefringence. Generically there will be some backgrounds for which one or both of these modes has a superluminal group velocity ($v_g>1$) implying a violation of causality. Only the special subset of NLED theories for which this cannot happen are causal. 

A standard causality condition is convexity of the Hamiltonian density 
$\HH$ as a function of ${\bf D}$.  This condition is intuitively obvious because whenever it is violated the Hessian of $\HH$ has a negative eigenvalue for some value of ${\bf D}$, which implies that linearisation is now about a saddle point of the energy density. If $\HH$ satisfies this convexity condition then the Lagrangian density $\LL$ will be a convex function of ${\bf E}$. 

Necessary and sufficient causality conditions were found in 2016 by Schellstede et al.\cite{Schellstede:2016zue}. These conditions are a collection of PDEs that we verified in \cite{Russo:2024kto} using the dispersion relation results of \cite{Bialynicki-Birula:1984daz,Russo:2022qvz}. They can be divided into two groups: weak-field causality conditions (which had previously\footnote{Previous to \cite{Russo:2024kto} but not to \cite{Schellstede:2016zue} which had (undeservedly) gone almost unoticed.} been shown to be equivalent to convexity of $\LL$ as a function of ${\bf E}$ \cite{Bandos:2021rqy}) and a novel 
strong-field causality condition that we showed in \cite{Russo:2025wph} to be equivalent to the following very simple inequality:
\be\label{SFCC}
\LL_U + 2U\LL_{UU}<0\, . 
\ee 
Some intuition into this condition is provided by the fact that it implies the Strong Energy Condition (SEC) for the stress-energy tensor \cite{Russo:2024xnh}, but it is a stronger condition than the SEC. The original Born model, for example, is acausal \cite{Schellstede:2016zue} even though its stress-energy tensor satisfies the SEC. Many other NLED models proposed or discussed in the literature for phenomenological purposes also fail to satisfy the strong-field causality condition \cite{Russo:2024kto}. Although it is not easily understood in simple terms, it is not difficult to see why there {\sl is} a strong-field causality condition, as we now explain. 

The first step is to restrict attention to {\sl static} backgrounds; little generality is lost because almost all stationary backgrounds are a Lorentz boost of some static background. The electric and magnetic background fields are now (anti)parallel and they pick out a `preferred' direction. The wave 3-vector ${\bf k}$ of some plane-wave perturbation of this background can be decomposed into its parallel and perpendicular components $(k_\parallel, {\bf k}_\perp)$. The two dispersion relations relating the angular frequency $\omega$ to ${\bf k}$ (which we label by $\pm$) now take the very simple form \cite{Bialynicki-Birula:1992rcm,Russo:2022qvz}
\be
\omega^2 = A_\pm |{\bf k}_\perp|^2 + k_\parallel^2\, ,  
\ee
where $A_\pm$ are functions of first and second partial derivatives of 
$\LL$ \cite{Bialynicki-Birula:1984daz,Russo:2022qvz} or of 
$\HH$ \cite{Mezincescu:2023zny}. 
The associated group velocities are
\be
v_g^\pm = \sqrt{\frac{A_\pm^2 |{\bf k}_\perp|^2 + k_\parallel^2}{A_\pm |{\bf k}_\perp|^2 + k_\parallel^2}}\, . 
\ee
Notice that $v_g^\pm =1$ for $|{\bf k}_\perp|=0$. 
If we require $v_g^\pm\le 1$ for all values of $A_\pm$ for which the group velocities $v_g^\pm$ are defined then the coefficients $A_\pm$ must satisfy the inequalities 
\be\label{AAA}
0 \le A_\pm \le 1\, . 
\ee
Some important aspects of causality are already visible in this formula.
To begin with, notice that $v_g^\pm =1$ for {\sl all} choices of ${\bf k}$ when $A_\pm=1$, which is the case for Maxwell, and hence $A_\pm \approx1$ in the weak-field limit for all NLED theories that have Maxwell as a free-field limit. More generally, either $A_+=1$ or $A_-=1$ for any conformal weak-field limit \cite{Denisov:2017qou,Bandos:2020jsw,Russo:2022qvz}. Thus, for generic NLED theories it is possible that one of $A_+$ or $A_-$ (or both) could be positive for a weak-field background. This possibility is what the weak-field causality conditions exclude. For NLED theories satisfying these weak-field causality conditions either $A_+$ or $A_-$, or both,  typically decrease as the energy density of the background fields increases, with one of them becoming negative for sufficiently strong and generic background fields. This possibility is what the strong-field causality condition excludes. 

We now focus on the special case of self-dual NLED theories. In this case,
using the general formula for $A_\pm$ derived in \cite{Bialynicki-Birula:1984daz,Russo:2022qvz}, and computing
the first and second partial derivatives of $\LL(U,V)$ using \eqref{CH-Lag}, one finds
(after a long calculation that we omit) the following remarkably simple results:
\be\label{magic}
A_- = \frac{1}{\dot\ell(\tau)^2}\, , \qquad 
A_+ = \frac{\dot\ell(\tau)}{\dot\ell(\tau)+2 \tau \ddot\ell(\tau)}\, . 
\ee
Recalling that $\dot\ell >0$,
the causality condition $A_-\leq 1$ reduces to $\dot\ell \geq 1$. 
The other weak field causality condition $A_+\leq 1$ now requires $\ddot \ell \geq 0$.
Therefore, from the weak-field causality condition {\sl alone}, we have deduced 
the causality conditions \eqref{causcon}. These conditions imply strong-field causality because they imply $A_\pm >0$. 

The  conditions \eqref{causcon}  were originally found in \cite{Russo:2025wph}.
but the above derivation of them is simpler in some respects, and it 
confirms the conclusion of  \cite{Russo:2025wph} that for self-dual theories the weak-field causality conditions imply the strong field causality conditions. This is one 
motivation for the study of self-dual NLED because many (if not most) NLED models (other than BI) that have been proposed in the literature for phenomenological purposes are acausal \cite{Russo:2024kto}, but they can be replaced by a self-dual model with the same Lagrangian for zero magnetic field, which is causal provided only that the weak-field convexity/causality conditions are satisfied \cite{Russo:2025wph}. The replacement for Born's original model, for example, is BI. 

We have focused above on the expression for $A_\pm$ in terms of the CH-function 
$\ell(\tau)$, but these expressions can be re-written in terms of $y$ and $\Omega(y)$ by using the relations \eqref{ell-Om}; one finds that  
\be\label{Apm-Om}
A_- =\frac{1}{y^2}\, , \qquad A_+ =\frac{y\Omega''}{y\Omega'' +2\Omega'}\, .
\ee
 Using these expressions for $A_\pm$ we may now re-examine the implications of \eqref{AAA}:
\begin{itemize}

    \item $A_->0$ but $A_-\le 1$  requires $\boxed{y\geq 1}$.

    \item Since $\Omega'\ge0$ from \eqref{Aux-y}, the condition $A_+\le1 $ implies that $y\Omega''+ 2\Omega'>0$,  and then $A_+>0$ implies $\boxed{\Omega''(y)> 0}$.
    
\end{itemize}
The boxed equations are precisely the causality conditions of \eqref{causalO}. 
Although we used the strong-field causality condition $A_+>0$, this is a consequence of $A_+\le1$ for self-dual theories with a weak-field expansion. The proof is as follows. Rewrite the expression for $A_+$ as 
\be
A_+ = \frac{1}{1+ \epsilon} \, , \qquad \epsilon = \frac{2\Omega'}{y\Omega''}\, . 
\ee
In the vacuum $\epsilon$ is zero, so it is small for weak fields and the inequality 
$A_+\leq 1$ holds if $\epsilon>0$. On any trajectory through the space of background fields from weak to strong, $\epsilon$ will change continuously but must remain positive because the only other way that $A_+\le1$ can be satisfied is if $\epsilon\leq -1$, 
and any continuous change from $\epsilon>0$ to $\epsilon<-1$ must pass through the interval $\epsilon\in (-1,0)$ in which $A_+>1$. Therefore, it is enough to impose only
$A_\pm \le1$: weak-field causality implies strong-field causality for self-dual NLED theories with a weak field expansion. 

The analysis in terms of $W$ is essentially the same. In this case we have 
\be
A_- = e^{-2\phi}\, , \qquad A_+ = \frac{W''(\phi)- W'(\phi)}{W''(\phi)+ W'(\phi)}\, ,  
\ee
and the conditions $0< A_\pm \le1$ imply the causality conditions of \eqref{causal1}; i.e. $W''(\phi)\ge W'(\phi)$ for $\phi\ge \phi_0$ (the vacuum value of 
$\phi$, for which $W'=0$) and $\phi_0\ge0$.

\subsection{The case of constant $A_-/A_+$}\label{subsec:MMB}

Recall that $A_+\ne A_-$ generically. If one views the waves as perturbations 
of a background that provides a static homogeneous optical medium for these waves
then this medium is birefringent whenever $A_+\ne A_-$. Born-Infeld is the exceptional case
with $A_+=A_-$. One could view as quasi-exceptional the case with 
\be\label{Apm-ratio}
\frac{A_-}{A_+} = e^{-2\gamma}\, , 
\ee
for some dimensionless constant $\gamma$, because this is the maximum of the ratio of the squares of the group velocities, and if $\gamma\approx 0$ the birefringence will be small irrespective of the strength of the background fields. 

For self-dual NLED we may use \eqref{Apm-Om} to deduce that \eqref{Apm-ratio} is equivalent to the second-order ODE
\be
\left[\sinh(\phi-\gamma)\right] W''(\phi) = \left[\cosh(\phi-\gamma)\right] W'(\phi),  
\ee
The general solution is 
\be
W(\phi) = T\cosh(\phi-\gamma) + W_0\, , 
\ee
for arbitrary constant $W_0$ and positive constant $T$. This is the MMB case that we have 
used as an illustrative example in earlier sections. For $\gamma=0$, and hence $A_+=A_-$ or ``no birefringence'', it is Born-Infeld. This is consistent with the old and famous result of Boillat and Plebanski that BI is the unique NLED theory (with a weak-field limit) that is ``exceptional'' in this regard. Although the derivation of this result here is limited to the class of self-dual NLED theories, we have extended it to show that the MMB theory has a similar ``quasi-exceptional'' status. For any other self-dual NLED the ratio $A_+/A_-$ will typically become large for strong fields.

\section{Analyticity and $\phi$-parity}\label{sec:phi-parity}
\setcounter{equation}{0}

In the Introduction we pointed out that an implication of our
auxiliary-field formulation of the general self-dual NLED is that the 
Lagrangian density $\LL(S,P)$ found by elimination of the auxiliary field 
$\phi$ from \eqref{bbb1} has a weak-field expansion in powers of $S$ and 
$\sqrt{S^2+P^2}$, but that there are cases (such as BI) for which this expansion is actually an ``analytic'' expansion in $S$ and $P^2$. We claimed that the necessary and sufficient condition for the expansion to be analytic (in this sense) is 
\be\label{Weven}
W(\phi)= W(-\phi)\,  .
\ee
We now explain why this is true. 

Let us first define $\widehat W(\phi) := W(-\phi)$ to be the ``$\phi$-parity dual'' of $W(\phi)$. Using this function in \eqref{bbb1} gives us the Lagrangian density defining the $\phi$-parity dual to the self-dual NLED defined by \eqref{bbb1}:
\be\label{hatL}
\widehat\LL(S,P;\phi) = (\cosh\phi) S + (\sinh\phi) \sqrt{S^2+P^2} - \widehat W(\phi)\, . 
\ee 
In general, $\widehat W\ne W$, so we have a pair of self-dual NLED theories,
paired by $\phi$-parity. By the {\sl field redefinition} $\phi\to -\phi$, we can rewrite \eqref{hatL} as 
\be
\widehat\LL(S,P;\phi) = (\cosh\phi) S - (\sinh\phi) \sqrt{S^2+P^2} - W(\phi)\, , 
\ee 
where we have used the fact that $\widehat W(-\phi) = W(\phi)$. 
Comparing this with \eqref{bbb1} we see that the only difference is the change of sign of the $\sqrt{S^2+P^2}$ term. Therefore, elimination of $\phi$ from $\LL$ and $\widehat\LL$
yields a pair of Lagrangian densities $\LL(S,P)$ and $\widehat\LL(S,P)$ with weak-field expansions in powers of $S$ and $\sqrt{S^2+P^2}$ that differ 
only by the sign of $\sqrt{S^2+P^2}$.

Suppose now that $\widehat W =W$. Then $\widehat\LL = \LL$, which implies that the weak-field expansion is in powers of $S$ and {\sl even} powers 
of $\sqrt{S^2+P^2}$, because this is the only way that the sign differences can cancel. In this case the expansion is actually in powers of $S$ and $P^2$. For the generic case with $\widehat W \ne W$, and hence $\widehat\LL\ne \LL$, not all sign differences can cancel, and the expansion must contain some odd power of $\sqrt{S^2+P^2}$. Thus, {\sl the condition \eqref{Weven} is both necessary and sufficient for an analytic weak-field expansion}. 

To summarize: within the class of self-dual Lorentz invariant NLED theories with weak-field expansions, there is a subclass for which this expansion is `analytic', and within the context of our new auxiliary field formulation this subclass is characterised by a  $\phi$-parity invariant potential function $W(\phi)$; i.e. one that is an even function of $\phi$. Born-Infeld is the simplest example. Another example of a self-dual NLED that is also $\phi$-parity self-dual and causal was found in \cite{Russo:2024ptw}; its potential function is
\be
W(\phi) = T\left[\cosh\phi  + a (\cosh\phi)^2 - (1+a)\right]\qquad (a>0). 
\ee

Self-dual NLED theories that are {\sl not} $\phi$-parity self-dual are members of $\phi$-parity doublets. The vacuum values of $\phi$ have opposite signs 
for such doublets:
\be 
\widehat\phi_0 = - \phi_0\, . 
\ee
This implies that at least one member of the $\phi$-parity pair is acausal unless $\phi_0=0$. For example, $\phi_0=\gamma>0$ for MMB;  its $\phi$-parity dual has $\gamma<0$ and is acausal. This remains true for all theories with ModMax as the weak-field limit because $\phi$-parity flips the sign of 
$\gamma$.  In contrast, any self-dual NLED with Maxwell as the weak-field limit has $\phi_0=0$. This can be seen from the expansion of $W(\phi)$ in \eqref{Wexpand} because this simplifies, for a Maxwell weak-field limit, to 
\be
W(\phi) = \sum_{n=2}^\infty c_n \phi^n\, ,  
\ee
for constant coefficients $c_n$, and in these cases $W'(0)=0$. However $\phi_0=0$ does not imply $\phi$-parity self-duality because the
above function $W(\phi)$ is $\phi$-parity invariant only if $c_{2k+1}=0$ for all integers $k\ge1$. We therefore expect there to be many 
$\phi$-duality pairs with $\phi_0=0$, and we also expect to find many pairs where one member of the pair is causal (by adjusting parameters). 
What is less clear is whether {\sl both} members of a $\phi$-parity dual pair can be causal. 

In fact, one may ask how it is possible for two self-dual NLED theories related by $\phi\to -\phi$ to be causal when $\phi\ge0$ is required for causality! Here it is especially important to appreciate that the causality relation $\phi\ge0$ is a restriction on the function $\phi(U,V)$ obtained by solving the auxiliary-field equation for $\phi$. These functions are different for the two members of a $\phi$-parity pair, 
and hence it is possible, in principle, for both of them to satisfy the $\phi\ge0$ 
causality condition (provided that $\phi_0=0$).  As the example of the following subsection 
demonstrates, such  causal $\phi$-parity pairs exist.

\subsection{A causal $\phi$-parity pair}

For what follows it is simpler to use the formulation in terms of the convex function 
$\Omega(y)$, with $y \to 1/y$ as the $\phi$-parity transformation. Consider the following $\Omega$ function\footnote{This yields a special case of a family of
self-dual NLED discussed in \cite{Russo:2024ptw}.}
\be
\Omega(y)= \frac{T}{4}\left( 3y+ y^{-3}\right) -T  \, . 
\ee
Its first and second derivatives are 
\be
\Omega'(y) = \frac{3T}{4} \left[ 1- y^{-4}\right]\, , \qquad \Omega''(y) = \frac{3T}{y^5} >0\, . 
\ee
The $y$ field equation \eqref{Aux-y} is a quadratic equation for $y^2$ but there is a unique solution for real positive $y$:
\be\label{y1}
y^{-1} = \sqrt{\Delta -\frac{2U}{3T}}\, , \qquad 
\Delta = \sqrt{1 - \frac{4V}{3T} + \left(\frac{2U}{3T}\right)^2}\, .   
\ee
Notice that $y\ge1$ because $y=1$ only for $U=V=0$ and it is easily verified that $y>1$ for weak fields. It is also easily checked that $\Omega''(y)>0$ for $y\ge1$. The causality conditions of \eqref{causalO} are therefore satisfied. The Lagrangian density is found by substitution for $y$ in \eqref{LUVy}; this yields 
\be\label{q3/4}
\LL(U,V) = T \left\{ 1- \left[\Delta + \left(\frac{4U}{3T}\right)\right] 
\sqrt{\Delta - \left(\frac{2U}{3T}\right)}\right\}\, .
\ee 
Notice that reality of $\LL$ requires $4V<3T$, which is an upper-bound
on $V$ that is already needed in \eqref{y1} for reality of $y$. The weak-field expansion is
\be\label{wfexp}
\LL = S + \frac{(S^2+P^2)}{6T} +\frac{\left(S^2 + P^2\right) \left(3S+2 \sqrt{S^2+ P^2}\right)}{54 T^2} +O(T^{-3}) \, , 
\ee
which confirms that the weak-field limit is Maxwell but that non-analyticity appears at the $S^3$ order.  

The $\phi$-parity dual of this causal self-dual NLED  has the $\Omega$-function
\be
\widehat\Omega(y) = \frac{T}{4}\left(3y^{-1} + y^3\right) -T\, .
\ee
The first and second derivatives are
\be
 \widehat\Omega'(y) = \frac{3T}{4}\left[y^2 - y^{-2}\right]\, , \qquad \widehat\Omega''(y) = \frac{3T}{2} \left[ y+ \frac{1}{y^3}\right] >0\, . 
\ee
The $y$ field equation is again a quadratic equation in $y$ with a unique solution for real positive $y$:
\be\label{y2}
y = \sqrt{ \frac{2V}{3T} + \widehat\Delta} \, , \qquad 
\widehat\Delta = \sqrt{1+ \frac{4U}{3T} + \left(\frac{2V}{3T}\right)^2}\, . 
\ee
Again we find that $y=1$ only for $U=V=0$, and that $y>1$ for weak fields, so $y\ge1$.  
It is also easy to verify that $\widehat\Omega''(y)>0$ for $y\ge1$, so the causality conditions of \eqref{causalO} are again satisfied. The dual Lagrangian density $\widehat\LL(U,V)$ can again be found by substitution for $y$ in \eqref{LUVy}, with $\Omega(y)$ replaced by $\widehat\Omega(y)$. The result is  
\be
\widehat\LL(U,V) = T \left\{ 1- \left[ \widehat\Delta -\frac{4V}{3T} \right] 
\sqrt{ \left(\widehat \Delta  + \frac{2V}{3T}\right) }\right\}\, .
\ee 
This can also be found directly from \eqref{q3/4} by making use of the fact that the expression for $y$ in \eqref{y1} and \eqref{y2} are related by
\be
y \ \leftrightarrow \ y^{-1}\, , \qquad (U,V) \  \leftrightarrow \ -(V,U)\, . 
\ee
Since $\widehat\Omega(y) = \Omega(y^{-1})$ by definition, this implies that 
$\widehat\LL(U,V) = \LL(-V,-U)$, as expected because this is the result of a $\phi$-parity transformation after elimination of the auxiliary field. The weak-field expansions of $\widehat \LL$ is the same as \eqref{wfexp} except for the sign of the $\sqrt{S^2+P^2}$ term at the $S^3$ order. 

\section{Simplified `analytic' self-dual NLED}\label{sec:RTgen}

Within the `analytic' (or $\phi$-parity self-dual) class of self-dual NLED a further remarkable simplification 
is possible, as we now explain. 

Consider the following Lagrangian density (inspired by a proposal in \cite{Lechner:2022qhb} that we briefly discuss below):
\be\label{L-vart}
\mathcal{L}(S,P; \phi, \vartheta) = \lambda(\phi,\vartheta) S + \mu(\phi,\vartheta) P - W(\phi)\, , 
\ee
where $\vartheta$ is a new angular auxiliary field ($\vartheta \sim \vartheta +2\pi$) and the coefficient 
functions $(\lambda,\mu)$ are 
\be\label{RT1}
\lambda = \cosh\phi  + \sinh\phi \cos\vartheta\, , 
\qquad \mu = \sinh\phi\sin\vartheta\, .
\ee
Notice that (spacetime) parity is preserved if we assign odd parity to the new 
auxiliary field $\vartheta$. Its field equation implies that 
\be
\cos\vartheta = \pm \frac{S}{\sqrt{S^2+P^2}} \, , \qquad \sin\vartheta = \pm \frac{P}{\sqrt{S^2+P^2}}\, ,  
\ee
and back-substitution yields 
\be
\LL = (\cosh\phi) S  \pm (\sinh\phi) \sqrt{S^2+P^2} - W(\phi) \, .
\ee
Without the potential $W$ the sign ambiguity would be physically irrelevant because it can be removed by the {\sl field redefinition} $\phi\to \pm \phi$. However, this field redefinition takes the above Lagrangian density, with the potential term, to 
\be
\LL_\pm  = (\cosh\phi) S + (\sinh\phi) \sqrt{S^2+P^2} - W(\pm \phi) \, . 
\ee
We see from this that the sign ambiguity is generically physically relevant because 
the sign distinguishes between one self-dual NLED and its $\phi$-parity dual. Generically, therefore, the Lagrangian density of \eqref{L-vart} fails to define a specific theory. However, for the special class
of analytic ($\phi$-parity self-dual) self-dual NLED, for which $W(-\phi)=W(\phi)$, it {\sl does}
define a definite theory. In fact, the Lagrangian density of \eqref{L-vart} gives us an equivalent Lagrangian density to \eqref{bbb1}, but one that is now {\sl linear} in $(S,P)$. 

We should mention here the similar idea proposed in \cite{Lechner:2022qhb} for an auxiliary-field formulation of the ModMax Lagrangian density linear in $(S,P)$. In that case there was no potential $W$ and our auxiliary field $\phi$ was the ModMax coupling constant $\gamma$, so the sign ambiguity led to two different NLED theories within the ModMax family. This problem was solved by the introduction of auxiliary fields, but their construction is not useful here because it would impose $\phi\ge0$ as a constraint that follows from an another auxiliary-field equation rather than from the $\phi$ field equation. The difference is important because (as already stressed above) the causality condition $\phi\ge0$ is not a condition on the independent unconstrained field $\phi$ but rather  on the function $\phi(S,P)$ found by solving the $\phi$ field equation.

To summarise: The Lagrangian density of \eqref{L-vart} defines a self-dual NLED provided 
$W(\phi)$ is $\phi$-parity invariant. Elimination of the auxiliary field $\phi$ yields the Lagrangian density $\LL(S,P)$ of an `analytic' self-dual NLED with a weak-field expansion in powers of $S$ and $P$, but the powers of $P$ will be {\sl even} powers because of (spacetime) parity invariance.

A Lagrangian density for BI that is linear in $(S,P)$ and involves two auxiliary fields was proposed many years ago by Ro{\v c}ek and Tseytlin \cite{Rocek:1997hi}. As we now show, their result is a special case of \eqref{L-vart}. We start from the observation that the functions $(\lambda,\mu)$ defined in \eqref{RT1} satisfy the following 
identity:
\be\label{RT-id}
\frac12 \left[ \lambda + \frac{1+\mu^2}{\lambda}\right] \ \equiv \ \cosh\phi\, .
\ee
Since $W(\phi)= T\cosh\phi$ for BI, we can rewrite \eqref{L-vart} {\sl for this case} as
\be
\LL(S,P; \lambda,\mu) = \lambda S + \mu P - \frac{T}{2}\left[ \lambda + \frac{1+\mu^2}{\lambda}\right]\, . 
\ee
Although $(\lambda,\mu)$ are here functions of $(\phi,\vartheta)$ we may take them to be the independent auxiliary fields, with $(\phi,\vartheta)$ defined implicitly by \eqref{RT1}. We then have precisely (but in 
a different notation) the BI Lagrangian density of \cite{Rocek:1997hi}. 

Returning to \eqref{L-vart}, we observe that it would be  natural to replace $(\phi,\, \vartheta)$ by the  complex auxiliary field 
\be
z= (\sinh\phi) e^{i\vartheta}\ .
\ee
The Lagrangian density of \eqref{L-vart} then takes the form
\be
\LL = \left[\Re(z) +\sqrt{1+z\bar z}\right] S + \Im(z) P -\mathcal{W} (z\bar z)\ , 
\ee
where $\mathcal{W}(\sinh^2\!\phi) = W(\phi)$. Born-Infeld corresponds to the
choice $\mathcal{W}=T\sqrt{1+z\bar z}$. Whether this observation has any utility remains to be seen.

\subsection{The Hamiltonian}

We have seen above that the Lagrangian density for any `analytic'  self-dual NLED may be written  as
\be
\LL(S,P;\phi) =  \lambda(\phi,\vartheta) S + \mu(\phi, \vartheta) P - W(\phi) \, , 
\ee
where $(\phi,\vartheta)$ are auxiliary fields and the coefficient functions $(\lambda,\mu)$ 
are the functions of them given in \eqref{RT1}.  To find the corresponding Hamiltonian density we first define (assuming cartesian space coordinates)
\be
{\bf D}  := \frac{\partial \LL}{\partial {\bf E}} = \lambda {\bf E} + \mu {\bf B}\, .
\ee
Solving for ${\bf E}$ we find that $\HH=D^iE_i - \LL$ becomes a linear function of the three independent rotation scalars 
\be
s = \frac{1}{2}\left(|D|^2 + |B|^2\right)\, , \quad \xi = \frac{1}{2} \left(|D|^2 - |B|^2\right)\, , \quad \zeta= {\bf D} \cdot {\bf B}\, .
\ee
Specifically, 
\be
\HH = (\cosh\phi) s - (\sinh\phi) \left(A \xi +B \zeta\right) + W(\phi)
\ee
where the identity \eqref{RT-id} has been used to simplify the coefficient of $s$, and
\be
A= \frac{\sinh\phi + \cosh\phi \cos\vartheta}{\cosh\phi + \sinh\phi\cos\vartheta}\, \qquad B= \frac{\sin\vartheta}{\cosh\phi + \sinh\phi\cos\vartheta}\, . 
\ee
Since $A^2+B^2 \equiv 1$, we can introduce a new angle $\theta$ such that
\be
A= \cos\theta\, , \qquad B= \sin\theta\, . 
\ee
Taking $(\phi,\theta)$ as a new basis for the auxiliary fields, we now have
\be\label{newangle}
\HH(s,\xi,\eta; \phi, \theta) =  (\cosh\phi) s - (\sinh\phi) \left[\xi \cos\theta + \zeta\sin\theta\right] + W(\phi)\, . 
\ee

As a check on this result, we observe that elimination of $\theta$ using the equation $\delta H/\delta\theta=0$  has the effect that 
\be
 (\sinh\phi)\left[\xi \cos\theta + \zeta\sin\theta\right]  \to \pm (\sinh\phi)\sqrt{\xi^2+\zeta^2} \, . 
 \ee
 The sign ambiguity is irrelevant  because the sign can be changed by the redefinition $\phi\to-\phi$, which leaves $W(\phi)$ unchanged by hypothesis (we are considering
only {\sl analytic} self-dual NLED). Choosing the upper sign, and using the fact that
  \be
  \xi^2+ \zeta^2 \equiv  s^2-p^2\, , \qquad p:= |{\bf D}\times {\bf B}|\, , 
  \ee
  we find that 
  \be
  \HH = (\cosh\phi) s - (\sinh\phi) \sqrt{s^2-p^2} + W(\phi)\, , 
  \ee
  which is just \eqref{ccc1} of the Introduction, in this case for a symmetric potential function $W(\phi)$.

\section{Energy conditions}\label{sec:energy}
\setcounter{equation}{0}

For any NLED theory with Lagragian density $\LL(S,P)$, the stress-energy tensor $\TT$ can be written in the form
\be\label{TT}
\TT_{\mu\nu} = \frac{A}{\sqrt{S^2+P^2}}\TT^{\rm Max}_{\mu\nu} - {\rm B} g_{\mu\nu}\, ,  
\ee
where $g$ is the metric on spacetime, which we take here to be the flat Minkowski spacetime, and 
$\TT^{\rm Max}$ is the Maxwell stress-energy tensor 
\be
\TT_{\mu\nu}^{\rm Max} = g^{\rho\sigma} F_{\mu\rho}F_{\nu\sigma} + S g_{\mu\nu}\, . 
\ee
The coefficients $({\rm A},{\rm B})$ are 
\be
{\rm A} = \sqrt{S^2+P^2}\LL_S\, , \qquad {\rm B} = S\LL_S + P\LL_P - \LL\, . 
\ee

The expression \eqref{TT} for the stress-energy tensor was used in \cite{Russo:2024xnh} to show that 
both the Dominant Energy Condition (DEC) and the Strong Energy Condition (SEC) are satisfied by any causal NLED, assuming zero vacuum energy, i.e. 
\be
\LL(0,0) =0\, . 
\ee
It can also be used to find expressions for the two Lorentz invariants that form a basis for scalar functions of the tensor $\TT$:
\be\label{T-inv}
\Theta := g^{\mu\nu} \TT_{\mu\nu}=-4{\rm B} \,  , \qquad 
\TT^2 :=g^{\mu\nu}g^{\rho\sigma} \TT_{\mu\rho}\TT_{\nu\sigma} = {\rm A}^2  +{\rm B}^2\, . 
\ee

Sufficient conditions for the DEC and SEC can found as simple inequalities on the $({\rm A},{\rm B})$ coefficients, as we now explain. 
\begin{itemize}

    \item DEC. This requires that $\xi^\mu\zeta^\nu \TT_{\mu\nu} \ge0$ for any two timelike vector fields 
    $(\xi,\zeta)$ with the same time orientation, which implies that $\xi\cdot \zeta <0$ (we use a ``mostly minus'' metric signature convection). Because the DEC is known to be satisfied for Maxwell, we see from  \eqref{TT} that a sufficient set of conditions for the DEC inequality to hold are ${\rm A}\ge0$ and ${\rm B}\ge0$.

    \item SEC. This requires that $\xi^\mu\xi^\nu[\TT - \tfrac12 \Theta g]_{\mu\nu}\ge0$ for any timelike vector field $\xi$, which we may normalise such that $\xi\cdot\xi=-1$. In a Lorentz frame for which $\xi= \partial_t$,  this inequality becomes 
    \be
    {\rm A} \left[\frac{\TT_{tt}^{\rm Max}}{\sqrt{S^2+P^2}}\right] - {\rm B} \ge 0
    \ee
    However, since 
    \be
    \TT_{tt}^{\rm Max} = \sqrt{S^2+P^2 + |{\bf E}\times {\bf B}|^2} \ge \sqrt{S^2+P^2}\, , 
    \ee 
    a sufficient condition for the SEC to be satisfied is ${\rm A}\ge {\rm B}$. 
\end{itemize}

The main aim of this section is to apply these general results to the specific case of self-dual NLED 
in the context of our new auxiliary field formulation. The stress-energy tensor may be found directly from the Lagrangian density \eqref{bbb1}; the $({\rm A},{\rm B})$ coefficients are 
\be\label{ABc}
{\rm A}= W'(\phi)\, , \qquad {\rm B}= W(\phi)\, ,  
\ee
where it should be understood that $\phi$ stands for the function $\phi(S,P)$ that solves the $\phi$ field equation. The invariants of \eqref{T-inv}  now take the very simple form 
\be
\Theta = -4W \, , \qquad  {\cal T}^2 = [W']^2 + W^2\, .  
\ee
It follows that $(W, [W']^2)$ is a basis for scalar functions of the stress-energy tensor for any self-dual NLED (excepting Maxwell/ModMax, but including BB electrodymamics, for which 
both $\Theta$ and $\mathcal{T}^2$ are zero). 

Of course, we already know that the causality conditions \eqref{causal1} must imply both the DEC and the SEC since this is true for generic NLED theories but it is instructive to see how this follows from the causality conditions on the potential function $W(\phi)$. To discuss energy conditions we should first fix the vacuum energy to be zero by choosing
\be
W(\phi_0)=0\, , 
\ee 
where $\phi_0$ is the vacuum value of $\phi$ defined by $W'(\phi)=0$. As explained in section \ref{sec:GC}, for causal theories $\phi_0$ is also the lowest possible value 
of $\phi$ solving the auxiliary-field equation. Moreover, since 
$W'>0$ for any $\phi>\phi_0$, it follows that $W(\phi)>0$ for $\phi>\phi_0$, and hence that both ${\rm A}$ and ${\rm B}$ are non-negative. This implies the DEC.

To see that the SEC also holds, we integrate the causality condition $W''-W'>0$ from $\phi_0$ to $\phi$ to find that 
\be
W'(\phi) - W(\phi)\ \ge\  W'(\phi_0) -W(\phi_0) =0 
\ee
and hence $W'>W$. From \eqref{ABc} we see that this is equivalent to ${\rm A}>{\rm B}$, which implies the SEC.

\section{Summary and Discussion}
\setcounter{equation}{0}

The aim of this paper has been to simplify many previous results in nonlinear electrodynamics, mainly 
to do with properties of self-dual theories, by providing simpler proofs of significant facts, 
improved physical intuition, and a new simple auxiliary-field formulation that facilitates 
these aims. A surprising feature of this formulation is that it `represents' the general non-conformal self-dual NLED as an extension of the conformal Maxwell/ModMax family rather than the general interacting self-dual NLED  as an 
extension of the free-field Maxwell theory.

The inspiration for this formulation was an ``alternative CH-construction'' of 
self-dual NLED that we introduced in previous work. The previous ``CH-construction'' is based 
on a solution to the self-duality PDE for the Lagrangian density $\LL$ in terms of a 
``CH-function'' $\ell$ that specifies the boundary conditions. Causality requires this 
function to be convex, and the conformal Maxwell/ModMax cases correspond to the convex linear 
functions. The ``alternative CH-construction'' expresses $\LL$ in terms of the Legendre 
transform of $\ell$, here called $\Omega$. This makes the Maxwell/ModMax cases special.  Whereas the Legendre transform of any nonlinear convex function $\ell(\tau)$ is another nonlinear convex function $\Omega(y)$, the Legendre transform of a {\sl linear} convex function  $\ell$ is a function $\Omega$ defined at a single point (and {\it vice-versa} because the Legendre transform is an involution for convex functions). This peculiarity is compensated by a new feature: the equation for $\Omega(y)$ determining $y$ (as a function of the electromagnetic fields) is the Euler-Lagrange equation for $\LL$ as a function of $y$,  and the convexity of $\Omega(y)$ ensures that the solution for $y$ is unique. This allows a re-interpretation of $y$ as an auxiliary field. 

The final step is to solve an inequality on $y$ to obtain an unconstrained  auxiliary-field $\phi$ with potential function $W(\phi) =\Omega(y),\, y=e^\phi$, as presented in the Introduction. We have provided details of this connection between the new auxiliary-field formulation and the previous ``CH-constructions'' because this is a simple way to establish generality and to find the causality conditions on the potential $W$. 

For generic NLED it is not a simple matter to determine the necessary and sufficient 
conditions for causality; they have been known for less than a decade, and we used them
more recently for self-dual NLED to find causality conditions on the CH-function 
$\ell$ mentioned above. The simplicity of the result suggests that a simpler derivation 
should be possible and it is, but the computations are still long. We have presented here
only the basic logic, with explanations of how causality can fail for both weak fields and strong fields, and how the causality conditions on $\ell$ can be deduced directly from dispersion relations; the same is true for $\Omega$, and hence $W$.

A significant feature of the new auxiliary-field formulation of self-dual NLED is that it makes manifest various special properties of self-dual NLED. One that is clearly seen from the Hamiltonian version is that the strong-field limit of {\sl any} self-dual NLED is the conformal ``BB electrodynamics'' found originally as a limit Born-Infeld; this is because the potential $W$ has dimensions of energy density and therefore sets the Born-scale, so the strong-field limit is $W\to0$, which yields the BB theory. 

Another property of self-dual NLED that is essentially manifest in the new formulation
is a simple relation between the Lagrangian and Hamiltonian, a ``substitution rule'' that allows one to be found from the other once both are expressed as functions of appropriate variables; it still depends on the result of a Legendre transform but
this variational problem is now the same for all self-dual NLED. 
We deduced this rule in earlier work but with much more effort. 

A related feature of the new formulation is that it makes manifest the fact that the weak-field expansion is an 
expansion in ``appropriate variables'' (either Lagrangian or Hamiltonian). 
For example, the fact that the Lagrangian density of \eqref{bbb1} is linear in Lorentz 
scalars $S$ and $\sqrt{S^2+P^2}$ tells us that the weak-field expansion will be a power series 
in these variables. This is significant because it is not always possible to rewrite such
a series as one in powers of $S$ and $P^2$ (odd powers of $P$ are excluded by parity). 

In earlier work we pointed out that the self-dual NLED with an ``analytic''  weak-field expansion (i.e. in powers of $S$ and $P^2$) are those for which (in the notation used here) $\Omega(y)$ is invariant under $y\to 1/y$. In terms of $W(\phi)$ this becomes invariance under $\phi\to-\phi$, which we have called $\phi$-parity. Thus  ``$\phi$-parity self-duality'' is equivalent to ``analyticity''. Generic NLED theories are not ``analytic'' and they arise in $\phi$-parity pairs; we have presented a simple explicit example with the property that both members of the pair are causal.  

For those theories that are ``analytic''  we might expect there to be an auxiliary-field formulation that makes manifest the fact that their weak-field expansions are power series in $S$ and $P^2$.  We have shown that this is true: for any ``analytic'' self-dual NLED it is possible to introduce a second (parity-odd and angle-valued) auxiliary field,  $\vartheta$, to linearise the Lagrangian density of \eqref{bbb1} in $S$ and $P$. For the choice of potential $W=T\cosh\phi$ it reduces to the Ro{\v c}ek-Tseytlin auxiliary-field formulation of Born-Infeld. We have thus found a generalisation of the Ro{\v c}ek-Tseytlin Lagrangian for BI to all causal `analytic' self-dual NLED. 

We conclude with some comments on possible generalisations and applications of our new auxiliary-field formulation for self-dual NLED. One obvious possibility is extensions to supersymmetric self-dual NLED \cite{Kuzenko:2000uh} as has been done using the Ivanov-Zupnik auxiliary-field formulation \cite{Kuzenko:2013gr,Ivanov:2013ppa}. Within our new auxiliary-field formulation, the natural starting point would be the general superconformal NLED (rather than the free-field NLED), which is supersymmetric ModMax \cite{Bandos:2021rqy,Kuzenko:2021cvx}. For the case of `analytic' self-dual NLED and the `linearised' Lagrangian density with two auxiliary fields, we have proposed a version of it with a single complex auxiliary field. Perhaps this version becomes natural in a supersymmetric context.

Another obvious generalisation would be to 6D chiral 2-form electrodynamics because of the one-to-one correspondence with 4D self-dual NLED \cite{Bandos:2020hgy}. This correspondence is clearest in the Hamiltonian formulation \cite{Russo:2025wph} but the Lagrangian version will likely apply to the Perry-Schwarz formulation \cite{Perry:1996mk} since there is strong evidence of its correspondence both to the Hamiltonian formulation \cite{Ferko:2024zth} and to the Lagrangian form of 4D self-dual NLED \cite{Bandos:2020hgy,Russo:2025wph}. 

Finally, another possible application of our auxiliary field formulation of self-dual NLED is to `$T\bar T$' flows since the construction of self-dual NLED in terms of the CH-function $\ell$ has already been applied in this context \cite{Babaei-Aghbolagh:2025cni}.



\section*{Acknowledgements}

JGR acknowledges financial support from grant 2021-SGR-249 (Generalitat de Catalunya) and  by the Spanish  MCIN/AEI/10.13039/501100011033 grant PID2022-126224NB-C21.





\end{document}